\documentclass[twocolumn,secnumarabic,amssymb, nobibnotes,prd,superscriptaddress]{revtex4}

\usepackage[utf8]{inputenc}
\usepackage{graphics,graphicx}
\usepackage{physics}
\usepackage{amsmath,amssymb,amsfonts,latexsym,cancel}
\usepackage{multirow}
\usepackage{hyperref}
\usepackage{bm}
\usepackage{xcolor}
\usepackage{ulem}
\usepackage{soul} 
\usepackage[capitalize]{cleveref}

\begin{document}

\newcommand{\quotes}[1]{``#1''}

\title{Rotating strange dwarfs and their indistinguishability from white dwarfs}

\author{Edson Otoniel}
\email{edson.otoniel@ufca.edu.br}
\affiliation{Instituto de Forma\c{c}\~ao de Educadores, Universidade Federal do Cariri, R. Oleg\'ario Emidio de Araujo, s/n – Aldeota, 63260-000 Brejo Santo, CE, Brazil}
\author{Jos\'e D. V. Arba\~nil}
\email{jose.arbanil@upn.pe}
\affiliation{Departamento de Ciencias, Universidad Privada del Norte, Avenida el Sol 461 San Juan de Lurigancho, 15434 Lima,  Peru}
\affiliation{Facultad de Ciencias F\'isicas, Universidad Nacional Mayor de San Marcos, Avenida Venezuela s/n Cercado de Lima, 15081 Lima,  Peru}

\author{Geanderson A. Carvalho}
\email{gacarvalho@utfpr.edu.br}
\affiliation{Departamento de F\'isica, Universidade Tecnológica Federal do Paraná, Medianeira, PR, Brazil}
\affiliation{Programa de P\'os-Gradua\c{c}\~ao em F\'isica e Astronomia, Universidade Tecnol\'ogica Federal do Paran\'a, Jardim das Americas, 82590-300, Curitiba, PR, Brazil}

\author{Fridolin Weber}
\affiliation{San Diego State University, Department of Physics, San Diego, 92182, California, USA}
\affiliation{University of California at San Diego, 9500 Gilman Drive, La Jolla, 92093, California, USA}

\date{\today}

\begin{abstract}
We investigate the structure of strange dwarfs, modeled as hybrid compact stars composed of a self bound strange quark matter core surrounded by a white dwarf like crust, within a fully relativistic framework. Static configurations are constructed by solving the Tolman Oppenheimer Volkoff equations, and uniformly rotating configurations are modeled within the Hartle–Thorne slow rotation expansion (to ${\cal O}(\Omega^2)$). We therefore interpret results at large fractional spins conservatively, and use the Kepler frequency mainly as a reference scale for comparing different masses and models. The stellar matter is described using a hybrid equation of state, in which the crust is modeled by a degenerate electron–ion system and the core by the MIT Bag Model. By comparing strange dwarfs with conventional white dwarfs across a range of rotation rates, we show that rotation inflates the radius and can reduce (in a quantifiable way) the separation between the two families in the $(M,R)$ plane, potentially masking structural signatures associated with the presence of a quark core. Our results highlight the importance of accounting for rotational effects when interpreting mass radius measurements and other global observables in the context of searches for exotic compact objects in current and future high precision surveys.
\end{abstract}

\maketitle

\section{\label{sec:level1} Introduction}

White dwarfs (WDs) are the most common final evolutionary stage of stars; in fact, about $90\%$ of the main-sequence stars will evolve into WDs \cite{shapiro_black_2008}. The most massive WDs have masses similar to that of the Sun and radii of about $10^3$ km, which makes them extremely compact, with an average mass density of $10^6$ g/cm$^3$ \cite{Caiazzo2021}. They can remain stable against gravitational collapse only because the electron degeneracy pressure counteracts gravity.

The baryonic matter that composes a WD is thought to be primarily carbon and oxygen, or sometimes neon and magnesium. In addition, spectroscopy measurements show that WD atmospheres are dominated by hydrogen or helium \cite{Bergeron2019}. However, according to the Bodmer-Witten hypothesis \cite{bodmer1971collapsed,witten1984}, strange quark matter (SQM) may be the absolute ground state of matter. If the SQM is indeed absolutely stable, then neutron stars, WDs, or even planets could possess a core of strange matter. This possibility opens a new perspective on the internal structure of WDs.

Motivated by this scenario, the possibility of WDs containing a SQM core, known as strange dwarfs (SDs), was explored decades ago \cite{Alcock1988Dec,Glendenning1995May}. Glendenning et al. \cite{Glendenning1995May} studied the stability of SDs under radial perturbations and showed that this class of objects can remain stable even when the normal matter density exceeds the neutron-drip density threshold, because the SQM core stabilizes the star. Subsequent works revisited the stability criteria of strange dwarf configurations using different mathematical approaches, see for instance \cite{Alford2017Sep}. At first sight, these studies might seem to lead to different conclusions regarding the stability of hybrid configurations. However, later analyses clarified that this apparent discrepancy arises from the physical assumptions adopted for the core--crust interface \cite{DiClemente2023Oct}. In particular, \cite{Pereira2018Jun,DiClemente2020May} showed that the radial stability of hybrid stars depends sensitively on the boundary conditions imposed at the phase transition, which are related to the dynamics of phase conversion across the density discontinuity. In this context, the interface may behave in different regimes, commonly referred to as slow or fast phase conversion, leading to distinct stability criteria.

Despite particular differences, SDs share several macroscopic similarities with ordinary WDs. For a given radius along the mass–radius relation, the SD is slightly less massive than its WD counterpart. On the other hand, for a fixed mass along the mass-radius relation, SDs have smaller radii compared to conventional WDs; see, for instance, \cite{DiClemente2024Aug}. Recently, in \cite{Perot2023May}, it is found that the tidal deformability of SDs deviated by about $8\%$ to near $50\%$, depending on the composition of the baryonic crust, from the values obtained for typical WDs. This makes tidal deformability a key parameter for distinguishing SDs from WDs through gravitational-wave astronomy. In \cite{Kurban2022Sep}, seven SD candidates are proposed by comparing their measured masses and radii with mass-radius diagrams. These candidates have smaller radii than other WDs with similar masses. The masses range from $0.02$ to $0.12M_{\odot}$, with radii between $9000$ and $15000$ km. This mass range is classified as that of extremely low-mass WDs \cite{Wang2022Aug}.

One important question concerning SDs is their origin. Current scenarios suggest two possible formation pathways: either a strange quark star accretes normal nuclear matter onto its surface, or a conventional WD captures clusters of SQM, known as strangelets \cite{DiClemente2024Aug}. Although the formation mechanism is highly relevant, it lies beyond the scope of this work. Here, we focus exclusively on investigating the macroscopic properties of rotating SDs.

The macroscopic structure, evolution, and observable properties of WDs are influenced by rotational effects. Rotation modifies the stellar equilibrium by providing additional centrifugal support, which increases the maximum mass a WD can sustain without collapsing and alters its mass radius relation \cite{Boshkayev2013}. Rapid rotation also produces measurable deviations from spherical symmetry, which in turn modify the moment of inertia, gravitational quadrupole moments, and tidal responses \cite{hartle1967,Becerra2019,Souza2020}; key parameters for interpreting binary evolution, gravitational-wave signals, and pulsation modes.

Rotation also plays an important role in the progenitors of Type Ia supernovae, where spin-up/spin-down processes impact explosion conditions and delay times \cite{DiStefano2011}. Observationally, WD rotation is inferred indirectly through rotational broadening of spectral lines, variability caused by magnetic spots, and asteroseismology; see, for example, \cite{daRosa2024}. Accurate theoretical modeling is therefore essential for correctly interpreting these observations. Thus, accounting for rotation is fundamental for realistic modeling of WDs or SDs and their potential role in astrophysics.

We organize this paper as follows. In Section~\ref{section2}, we present the theoretical framework adopted throughout this work, introducing the equations governing stellar equilibrium in general relativity and describing the Hartle--Thorne formalism used to model uniformly rotating compact stars within the slow-rotation approximation. This section also defines the physical assumptions and limits under which the rotational treatment remains valid. In Section~\ref{sec:level2}, we construct the hybrid equation of state employed to model strange dwarfs, detailing both the description of the hadronic crust, representative of white dwarf like matter, and the strange quark matter core modeled within the MIT Bag Model. We also explain the matching conditions at the core-crust interface and discuss the role of the electrostatic layer in ensuring mechanical equilibrium between the two regions. The numerical results are presented and discussed in Section~\ref{sec:results}. In this section, we analyze the relativistic stellar structure obtained from the adopted equation of state, considering both static and uniformly rotating configurations. Particular emphasis is placed on assessing the sensitivity of the stellar properties to variations of the bag constant and on quantifying the impact of rotation on global observables such as mass, radius, and compactness, allowing for a direct comparison between strange dwarfs and conventional white dwarfs. Finally, in Section~\ref{sec:conclusion}, we summarize the main findings of this study, discuss their physical implications within the context of compact star modeling, and outline the limitations of the present approach, as well as possible extensions and future directions motivated by observational and theoretical developments.

\section{Stellar equilibrium equations, Hartle-Thorne equations, and equation of state}\label{section2}

For the sake of completeness, we begin by presenting the Einstein field equations that govern a system containing matter, which, in geometric units, is given by
\begin{equation}\label{Einstein_equation}
    G_{\mu\nu}=8\pi T_{\mu\nu},
\end{equation}
with $\mu, \nu,$ etc. running from $0$ to $3$. $G_{\mu\nu}$ represents the Einstein tensor and $T_{\mu\nu}$ stands the energy-momentum tensor. For the fluid contained in the SD, the energy-momentum tensor is placed in the form
\begin{equation}\label{tem}
T_{\mu\nu} = (p+\rho)u_{\mu}u_{\nu}+pg_{\mu\nu},
\end{equation}
$p$ and $\rho$ represent the fluid pressure and the energy density, respectively. $u_{\mu}$ depicts the $4$-fluid velocity which follows condition $u_{\mu}u^{\mu}=-1$.

\subsection{Stellar equilibrium equations}\label{stellareq}

To describe the interior static spherically symmetric spacetime, the line element takes the form:
\begin{equation}\label{line_element}
ds^2 = -e^{\Phi(r)} dt^2 +e^{\Lambda(r)} dr^2 + r^2\left(d\theta^2 + \sin^2{\theta}d\phi^2\right),
\end{equation}
with $t, r, \theta,$ and $\phi$ representing the Schwarzschild coordinates. The functions $\Phi(r)$ and $\Lambda(r)$ depend only on the radial coordinate.

By means of the non-zero equations of the Einstein field equation \eqref{Einstein_equation} obtained by considering the energy momentum tensor for a perfect fluid \eqref{tem} and the previously defined metric \eqref{line_element} we find the relations:
\begin{eqnarray}
&&m' =4 \pi r^2 \rho,\label{eq_mass} \\
&&p' =- (\rho + p)\left(\frac{m+4 \pi r^3 p}{r^2}\right)e^{\Lambda}, \label{eq_p}\\
&&\Phi'=2\left(\frac{m+4 \pi r^3 p}{r^2}\right)e^{\Lambda}, \label{eq_g00}
\end{eqnarray}
where
\begin{equation}
e^{\Lambda}=\left(1-\frac{2m}{r}\right)^{-1}.
\end{equation}
The primes $(\,'\,)$ depict the derivative with respect to the radial coordinate $r$, and $m$ represents the mass contained within a radius $r$.

The stellar equilibrium equations, also known as the Tolman-Oppenheimer-Volkoff (TOV) equations, Eqs. \eqref{eq_mass}-\eqref{eq_g00}, Ref. \cite{tolman1939static,oppievolkoff}, are integrated from the center $r=0$ towards this star's surface $r=R$. The process starts considering $r=0$
\begin{eqnarray}
&&m(0)=0,\quad \rho(0)=\rho_c,\quad p(0)=p_c,\nonumber\\
&&\Lambda(0)=0,\quad {\rm and}\quad\Phi(0)=\Phi_c,
\end{eqnarray}
and it ends when in $r=R$
\begin{equation}
p(R) = 0.
\end{equation}
At the surface of the star, the interior and the exterior line elements match smoothly; thus, at this point, the potential metrics follow the relation:
\begin{equation}
e^{\Phi(R)}=\frac{1}{e^{\Lambda(R)}}=1 - \frac{2M}{R},
\end{equation}
with $M$ and $R$ being the mass and radius of the star, respectively.

\smallskip

\subsection{Hartle-Thorne equations}
The Hartle--Thorne formalism is a perturbative expansion in the stellar angular velocity, truncated at second order, and it is strictly controlled when the dimensionless rotation parameter (e.g., $\Omega^2 R^3/(GM)$) remains small. For this reason, throughout this work we use the Hartle--Thorne sequences to (i) quantify leading order rotational trends, and (ii) identify the onset of a potential observational degeneracy between SDs and WDs. When we quote fractions of the Kepler frequency, $\Omega/\Omega_K$, we employ $\Omega_K$ as a convenient normalization of the spin scale for each mass and EoS. However, configurations extremely close to mass shedding should ultimately be revisited with fully two dimensional relativistic rotating-star codes, and any statement of ``indistinguishability'' near $\Omega\sim\Omega_K$ should be interpreted in this conservative sense.

Following \cite{hartle1967,hartle1968}, to investigate the effects of rotation on the structure of compact stars, we used the spacetime metric with axial symmetry:
\begin{equation}\label{rotating_metric}
ds^2=-e^{2\nu}dt^2+e^{2\lambda}dr^2+e^{2\mu}d\theta^2+e^{2\psi}\left(d\phi-\omega dt\right)^2,
\end{equation}
with the potential metrics of the form:
\begin{eqnarray}
&&\hspace{-0.7cm}e^{2\nu}=e^{\Phi(r)}\left[1+2h_0(r)+2h_2(r)P_2(\cos\theta)\right],\\
&&\hspace{-0.7cm}e^{2\lambda}=e^{\Lambda(r)}\left[1+\frac{2e^{\Lambda(r)}}{r}\left[m_0(r)+m_2(r)P_2(\cos\theta)\right]\right],\\
&&\hspace{-0.7cm}e^{2\mu}=r^2\left[1+2k_2(r)P_2(\cos\theta)\right],\\
&&\hspace{-0.7cm}e^{2\psi}=r^2\sin^2\theta\left[1+2k_2(r)P_2(\cos\theta)\right].
\end{eqnarray}
In the line element \eqref{rotating_metric}, the terms $h_l(r)$, $m_l(r)$, and $k_l(r)$ correspond to the rotational perturbations of order $\Omega^2$, specifically the contribution of both the monopole ($l = 0$) and the quadrupole ($l = 2$); setting $k_0(r) = 0$ reflects the gauge choice adopted by Hartle \cite{hartle1967}. $P_l(\cos\theta)$ denotes the Legendre polynomial of degree $l$.

Meanwhile, the angular velocity of the local inertial frame $\omega(r)$ encapsulates the frame dragging effect induced by stellar rotation, emerging as a first order contribution in the expansion with respect to the star's angular velocity $\Omega$; with this latter running from $0$ to the Kepler frequency $\Omega_K$ \cite{glendening_weber1994,weber_glendening2012,weber_glendening1991}, which is determined by \cite{friedman_ipser1986}:
\begin{equation}
\Omega_K=\omega+\frac{\omega'}{2\psi'}+e^{\nu-\psi}\sqrt{\frac{\nu'}{\psi'}+\left(\frac{\omega'e^{\psi-\nu}}{2\psi'}\right)^2},
\end{equation}
with the primes indicating the partial derivative with respect to the radial coordinate. In the first order contribution specifically in the dipolar sector ($l = 1$) Hartle \cite{hartle1967} showed that $\omega(r)$, which describes the angular velocity of local inertial frames relative to a distant observer, satisfies a second-order differential equation derived from the $t\phi$ component of Einstein's field equations given by
\begin{equation}\label{rotational_equation}
    \frac{1}{r^4}\left(r^4j(r)\frac{d\omega(r)}{dr}\right)-\frac{4}{r}\frac{dj}{dr}(\Omega-\omega(r))=0,
\end{equation}
with $j(r)=e^{-(\Lambda+\Phi)/2}$. Equation \eqref{rotational_equation} arises directly from the $t~\phi$ component of Einstein’s field equations.

For a constant rotation ($\Omega$ constant), equation \eqref{rotational_equation} could be rewritten in the form:
\begin{equation}\label{rotational_equation1}
    \frac{1}{r^4}\left(r^4j(r)\frac{d{\bar\omega(r)}}{dr}\right)-\frac{4}{r}\frac{dj}{dr}{\bar\omega(r)}=0,
\end{equation}
where the new parameter follows ${\bar\omega}\equiv\Omega-\omega$. It represents the angular velocity of the fluid measured relative to the local frame of reference.

Equation \eqref{rotational_equation1} is solved from the center $r=0$ to the surface of the star $r=R$. This process begins by considering the conditions at the origin of the star:
\begin{equation}
    {\bar\omega}={\rm Constant}, \quad\quad \left(\frac{d{\bar\omega}}{dr}\right)_{r=0}=0.
\end{equation}
Outside the star ($r>R$), where $\rho=p=0$ and $j=1$, the next solution is obtained:
\begin{equation}
    {\bar\omega}=\Omega-2\frac{J}{r^3},
\end{equation}
with $J$ being the total angular momentum of the star. Due to the lack of dynamical degrees of freedom at the stellar surface, both ${\bar\omega}$ and its radial derivative $d{\bar\omega}/dr$ remain continuous at $r = R$. Consequently, at this boundary, the angular momentum $J$ and the angular velocity $\Omega$ are given, respectively, by the following expressions:
\begin{equation}
    J=\frac{R^4}{6}\left(\frac{d{\bar\omega}}{dr}\right)_{r=R},\quad\quad\Omega={\bar\omega}(R)+\frac{2J}{R^3}.
\end{equation}
Since there is flexibility in selecting the value of $\bar{\omega}$ at the stellar core, the resulting angular velocity $\Omega$ typically does not match a preferred target value. Thus, one rescale the function ${\bar\omega}$ as follows:
\begin{equation}
{\bar\omega}_{\rm new}={\bar\omega}_{\rm old}\frac{\Omega_{\rm new}}{\Omega_{\rm old}}.
\end{equation}

\section{\label{sec:level2} Equation of state}

To investigate rotating SDs, we consider a crust composed of a gas of degenerate electrons and a core of SQM. Additionally, we match the two phases at the neutron-drip point by imposing continuity of the (mechanical) fluid pressure, $p_{\rm crust}=p_{\rm core}=p_{\rm drip}$, while allowing for a discontinuity in the energy density across the interface, as expected for a first order transition and as in the standard strange dwarf construction \cite{Glendenning1995May}. In this idealized treatment, the thin electrostatic layer that ensures global charge neutrality and prevents immediate conversion of the crust (typical thickness $\ll 1\,$cm in physical units) is neglected in the macroscopic structure equations, so the interface is implemented as a sharp boundary at a single radius. We emphasize that the present matching prescription specifies the equilibrium branch of interest, but dynamical stability across a sharp interface depends on the phase conversion and boundary condition prescription (see, e.g., \cite{Alford2017Sep,Pereira2018Jun,DiClemente2020May}). 

This matching prescription corresponds to the slow phase-conversion regime at the interface. In this regime, matter does not convert instantaneously across the phase boundary during radial perturbations, which has important implications for the stability properties of hybrid configurations. This interpretation is consistent with the framework discussed in Refs.~\cite{Pereira2018Jun,DiClemente2020May}. These two fluid matter equations of state are described in the following subsections.

\subsection{\label{sec:level2_1} Crust Matter Composition}

In the seminal works \cite{salpeter_energy_1961,hamada_models_1961}, Salpeter and Hamada in $1961$ proposed that the composition of a WD matter is primarily made up of atomic nuclei immersed in a completely degenerate electron gas. In more recent work, Otoniel and collaborators in \cite{Otoniel_2019} carried out an approach where the equation of state (EoS) that describes the magnetic fluid contained in a WD is derived using updated atomic mass evaluations (see, for instance, Ref. \cite{wang_ame2012_2012,audi_ame2012_2012} and the references therein). Therefore, disregarding the magnetic field within the fluid, we consider that the pressure in SD crust is primarily due to degenerate electrons and the ionic lattice, such is assumed in the internal pressure of WDs \cite{shapiro_black_2008}. In this way, using this formalism, we regard that the total pressure in the crust of the SD is given by:
\begin{equation}\label{crust_matter_composition}
p_{\rm crust}\left(k_F\right)=p_L(Z)+\frac{1}{3 \pi^2 h^3} \int_0^{k_F} \frac{k^4}{\sqrt{k^2+m_e^2}} d k\,.
\end{equation}
The first term on the right hand side of equation \eqref{crust_matter_composition} depicts the pressure generated by the relativistic degenerate electron gas, which can be expressed as follows:
\begin{equation}
p_L(Z) = \frac{1}{3} C e^2 n_e^{4/3} Z^{2/3} \,.
\end{equation}
The last relation stems from Coulomb interactions between ions organized in a crystalline lattice, which is commonly assumed to adopt a body centered cubic (bcc) configuration within the interiors of WDs. In this equation, the dimensionless constant $C$ is determined by the geometry of the lattice and takes the value $-1.444$ for a bcc structure. The negative sign signifies that the electrostatic potential energy in the lattice corresponds to a binding interaction. The quantities $e^2$, $n_e$, and $Z$ correspond, respectively, to the strength of the Coulomb interaction between charged particles, the density of the electron number which, in fully ionized matter, is directly related to the density of positive ions as a result of charge neutrality and the atomic number of the element.

The second term on the right hand side of equation \eqref{crust_matter_composition} represents the relativistic dispersion relation integrand of the electrons, originating from the distribution of electron momenta $k$ up to Fermi momentum $k_F$, which determines the highest occupied momentum state at zero temperature, respectively. The symbol $m_e$ denotes the electron rest mass, incorporating relativistic effects into the pressure calculation, which is an essential aspect for accurately modeling high density environments. The factor $1/(3 \pi^2 h^3)$ arises from the proper normalization of the three dimensional momentum space volume, where $h$ denotes Planck’s constant. This coefficient guarantees that the integral over the distribution of electron momenta is dimensionally consistent and accurately scaled. When combined, these components yield the EoS for WD matter in the absence of magnetic fields, accounting for the influence of electron degeneracy and the structural role of the ionic lattice.

Following~\cite{chamel_stability_2013}, we assume that the total energy density $\rho_{\rm crust}(k_F)$ in the matter located in the crust of the strange dwarf which includes the contributions from nuclei, electrons, and the ionic lattice is expressed as
\begin{equation}
\begin{aligned}
\rho_{\rm crust}\left(k_F\right) & =\rho_L+\rho_e+\rho_i -\rho_\epsilon, \\
& =C e^2 n_e^{4/3} Z^{2/3}+\frac{1}{\pi^2 h^3} \int_0^{k_F} \sqrt{k^2+m_e^2} k^2 dk\\
&+n_i M(Z,A) - n_e m_e.
\end{aligned}
\end{equation}
As can be seen, the total energy density is composed of several distinct contributions. The first component, $\rho_L$, accounts for the energy density associated with the Coulomb lattice formed by the ions. The second, $\rho_e$, corresponds to the energy density of the degenerate electron gas, which includes the relativistic energy of electrons integrated from zero momentum up to the Fermi momentum. The third term, $\rho_i$, refers to the rest mass energy density of fully ionized atomic nuclei, where $n_i$ denotes the ion number density and $M(Z,A)$ is the nuclear mass of an ion with atomic number $Z$ and mass number $A$. The fourth contribution, $\rho_{\epsilon}$, serves as a correction by subtracting the electron rest mass energy that is already implicitly included in the nuclear mass $M(Z,A)$, thereby preventing the electron rest energy from being counted twice in the total. Taken together, these components form a self-consistent and physically complete expression for the crust energy density of SD matter under the assumptions of full degeneracy, absence of magnetic fields, and a crystallized plasma structure. In our study, we adopt carbon as the elemental composition of the crust material, setting $Z = 6$ and $A = 12$, which corresponds to fully ionized ${}^{12}\mathrm{C}$. The nuclear mass $M(6,12)$ used in the calculations is taken from experimental atomic mass evaluations, ensuring alignment with the most up-to-date empirical measurements \cite{wang_ame2012_2012,audi_ame2012_2012}.

Following the classical approach in \cite{glendenning_1995,benvenuto_structure_1996}, we therefore assume a pure carbon composition for the hadronic crust of the SD models. This choice provides both physical consistency and observational relevance, since most well studied WDs exhibit outer layers dominated by C/O mixtures. From a structural perspective, adopting heavier compositions such as O/Ne/Mg only shifts the mass radius curve slightly toward smaller radii, by less than $\sim3\%$ in both $M$ and $R$, as shown in relativistic calculations for massive WDs \cite{malheiro_relevance_2021}. Such deviations lie well within the current observational uncertainties \cite{mathews_analysis_2006}, and therefore do not affect the global trend that rotating SDs remain observationally indistinguishable from ordinary WDs. Preliminary calculations with O/Ne/Mg crusts confirm that this indistinguishability persists within the expected precision of future \textit{Gaia} and LISA observations. Hence, the adoption of a carbon crust represents a consistent and conservative approximation for the present investigation.

\subsection{Core Matter Composition}\label{sec:bag}

For the EoS employed in the core of the SD, we adopt one that describes quark matter with a self-bound phase composed of deconfined $u$, $d$, and $s$ quarks. In this framework, matter is assumed to be a zero temperature Fermi gas consisting of massless $u$, $d$, and $s$ quarks, enclosed by the QCD vacuum energy density, commonly referred to as the bag constant ${\cal B}$. Modifications arising from gluonic dynamics and color-superconducting pairing effects are omitted in this treatment, given that their influence on the structure of low mass configurations is expected to be minimal, typically below the $1\%$ level \citep{Alford1999}. Consequently, the resulting EoS takes the form:
\begin{equation}\label{eq:bag}
 p_{\rm core}=\frac{{\rho_{\rm core}}-4{\cal B}}{3},
\end{equation}
with $\rho_{\rm core}$ being the energy density of the quark. In line with the classical framework of
\cite{Alcock1986} and the phenomenological calibration of \cite{Farhi1984}, we employ a bag constant of ${\cal B^{\rm 1/4}}=\{135,145,160\}\,{\rm MeV}$. Following common practice, we quote the bag parameter through $B^{1/4}$ (in MeV) and convert it to $B$ in $\mathrm{MeV\,fm^{-3}}$ using $B = (B^{1/4})^{4}/(\hbar c)^{3}$. This choice ensures an energy per baryon in the range $\rho_{\rm core}/A = 875\text{–}894\;\mathrm{MeV}$, remaining well beneath the stability threshold set by iron at $930.4\;\mathrm{MeV}$, thereby satisfying the absolute stability condition for SQM.

\section{Results}\label{sec:results}

\subsection{Sensitivity of the hybrid EoS to the bag constant}\label{Eos}

\begin{table*}[ht]
\centering
\caption{Sensitivity of the maximum mass with the bag constant. For each value of ${\cal B}$, the maximum mass $M_{\max}$, the corresponding radius $R(M_{\max})$ and central density $\rho_c(M_{\max})$, and the relative variation $\Delta M_{\max}({\cal B})$ with respect to the reference value ${\cal B}_{\rm ref}^{1/4} = 145~\mathrm{MeV}$ are indicated.}
\label{tab:Bscan}
\begin{tabular}{lcccc}
\hline\hline
${\cal B}^{1/4}$ [MeV]
& $M_{\max}$ [$M_\odot$]
& $R(M_{\max})$ [km]
& $\rho_c(M_{\max})$ [$10^6\,\mathrm{g\,cm^{-3}}$]
& $\Delta M_{\max}$ [\%] \\
\hline
135 & 1.3596 & 447.15 & 173.234 & $-0.0370$ \\
145 & 1.3601 & 452.02 & 230.61 & $0.0000$ \\
160 & 1.3608 & 452.91 & 341.98 & $+0.0520$ \\
\hline\hline
\end{tabular}
\end{table*}

Throughout this section, we adopt the cold limit (\(T = 0\)), which is well justified for compact objects older than \(\gtrsim 10^{8}\) yr, for which thermal contributions to the pressure are negligible compared to electron degeneracy. In addition, SDs may retain residual magnetic fields inherited from their WD progenitors, typically in the range \(10^{3}\)–\(10^{5}\) G \citep{Baghdasaryan2016}. Such fields are dynamically irrelevant for global hydrostatic equilibrium: they do not significantly modify the mass radius relation, especially compared to the few percent variations induced by the bag constant scan summarized in Table~\ref{tab:Bscan} and quantified by \(\Delta M_{\max}(B)\) in Eq.~\eqref{eq:DeltaM}.

However, magnetically induced anisotropies can affect percent level features of the crustal structure and the directional components of the stress tensor, as shown by \cite{Rodriguez2023}. Although these effects remain subdominant in the context of equilibrium modeling, they may become relevant for oscillation spectra or multipolar deformations and thus may provide complementary diagnostics in future studies. A fully magnetostatic treatment, incorporating magnetic-field-dependent corrections to the EoS and anisotropic pressure contributions, will be addressed in a forthcoming work.

To check the sensitivity of the EoS with the change of the bag constant in the quark-matter sector, we considered three representative values of the bag constant (${\cal B^{\rm 1/4}}=\{135,145,160\}\,{\rm MeV}$), while keeping all other microphysics and the crustal composition fixed (pure \(^{12}\)C; same electrostatic matching and transition criterion). For each ${\cal B}$ considered, we construct the hybrid EoS and compute the corresponding sequence $M-R$ by integrating the TOV equations.

\begin{figure}[h]
    \centering
    \includegraphics[width=0.48\textwidth]{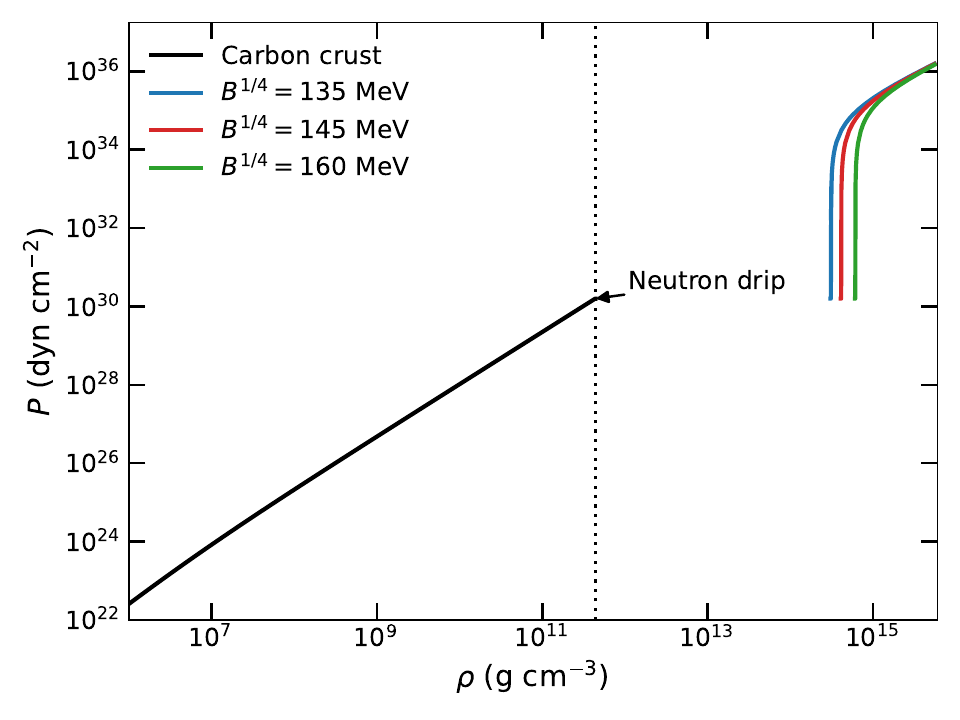}
    \caption{Fluid pressure against the energy density (carbon crust + SQM core) for ${\cal B^{\rm 1/4}}=\{135,145,160\}\,{\rm MeV}$. Differences appear only above neutron-drip density, where the SQM core starts to play a significant role. These trends explain the small shifts observed in the $M$-$R$ curves in Fig.~\ref{fig:mass_radiuas_bag}.}
    \label{fig:eos_multi_B}
\end{figure}

Figure~\ref{fig:eos_multi_B} shows the fluid pressure versus energy density profile for three different bag constants, highlighting the abrupt transition between the crust dominated by electron degeneracy pressure and the core, composed of quark matter. This discontinuity provides the microscopic foundation for the macroscopic stellar sequences discussed below. At low and intermediate densities, \(\rho \lesssim \rho_{\rm drip}\), the curves coincide because they are independent of the description of the quark matter. However, above the neutron-drip density, where the SQM core begins to play a significant role, smaller values of ({\cal B}) lead to higher fluid pressure at a fixed energy density, whereas larger values of ({\cal B}) result in lower fluid pressure. This behavior is fully consistent with the standard MIT bag-model phenomenology.

\begin{figure}[h]
    \centering
    \includegraphics[width=0.48\textwidth]{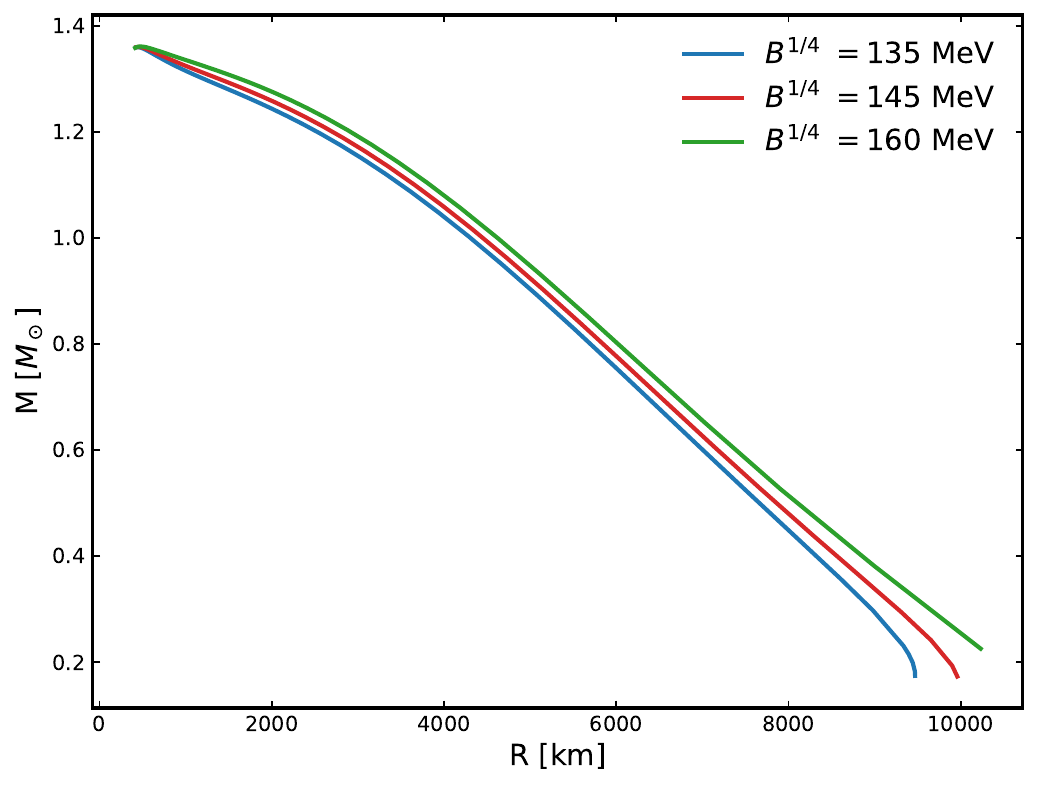}
    \caption{Mass-radius relations for SDs computed with ${\cal B^{\rm 1/4}}=\{135,145,160\}\,{\rm MeV}$.}
    \label{fig:mass_radiuas_bag}
\end{figure}

Once the EoS is defined, we solve the TOV equations to analyze the equilibrium of non-rotating stars for different values of ${\cal B}$. In this way, the effects of the bag constant on the mass, normalized in solar masses, as a function of the radius are shown in Fig.~\ref{fig:mass_radiuas_bag} considering three different values of ${\cal B}$. The overall shape of the $M-R$ curves is essentially preserved across the range of the bag constant considered. For each value of ${\cal B}$, we compute the maximum mass \(M_{\max}\), the corresponding radius \(R(M_{\max})\), and the central density \(\rho_c(M_{\max})\); see Table \ref{tab:Bscan}. To make the comparison explicit, we define the relative variation of the maximum mass with respect to the reference value \({\cal B}_{\rm ref}^{1/4} = 145~\mathrm{MeV}\),
\begin{equation}
  \Delta M_{\max}({\cal B}) \;=\;
  \frac{M_{\max}({\cal B}) - M_{\max}({\cal B}_{\rm ref})}{M_{\max}({\cal B}_{\rm ref})}
  \times 100\% \, ,
  \label{eq:DeltaM}
\end{equation}
which remains at the level of only a few percent over the explored range. This analysis shows that reasonable variations of the bag constant produce \(M\)-\(R\) curves that are nearly indistinguishable within current and near future measurement uncertainties.

\subsection{Effects of the rotation on SDs}

The static background is obtained by integrating Eqs.~\eqref{eq_mass}--\eqref{eq_g00} with a standard adaptive Runge-Kutta scheme, and the Hartle--Thorne perturbation equations are then integrated on top of the same background. 

The equilibrium properties of SDs were computed by solving the TOV equations for the static case and then extended for rotating configurations using the Hartle-Thorne slow rotation formalism. The adopted EoS, which depicts a crystalline carbon crust and a SQM core described by the MIT Bag Model, naturally introduces a density discontinuity at the core-crust interface. This discontinuity is a key feature that distinguishes SDs from ordinary WDs.

For the static sequences, we restrict attention to the usual stable branch selected by the turning-point condition along the one-parameter family of equilibria, $dM/d\rho_c>0$ (equivalently, before the first maximum of $M(\rho_c)$), which is a standard necessary criterion for radial stability of cold compact stars. For hybrid configurations with a sharp interface, full dynamical stability can depend on the assumed boundary conditions and phase-conversion rate at the interface; we therefore interpret $dM/d\rho_c>0$ as a practical screening criterion and refer the reader to \cite{Alford2017Sep,Pereira2018Jun,DiClemente2020May} for detailed treatments of radial modes in the presence of a density discontinuity. For rotating models within the Hartle--Thorne approximation, the same turning-point logic provides a useful guide, but configurations near $\Omega\sim\Omega_K$ should be assessed with fully two-dimensional methods.

\begin{figure}
    \centering
    \includegraphics[width=0.48\textwidth]{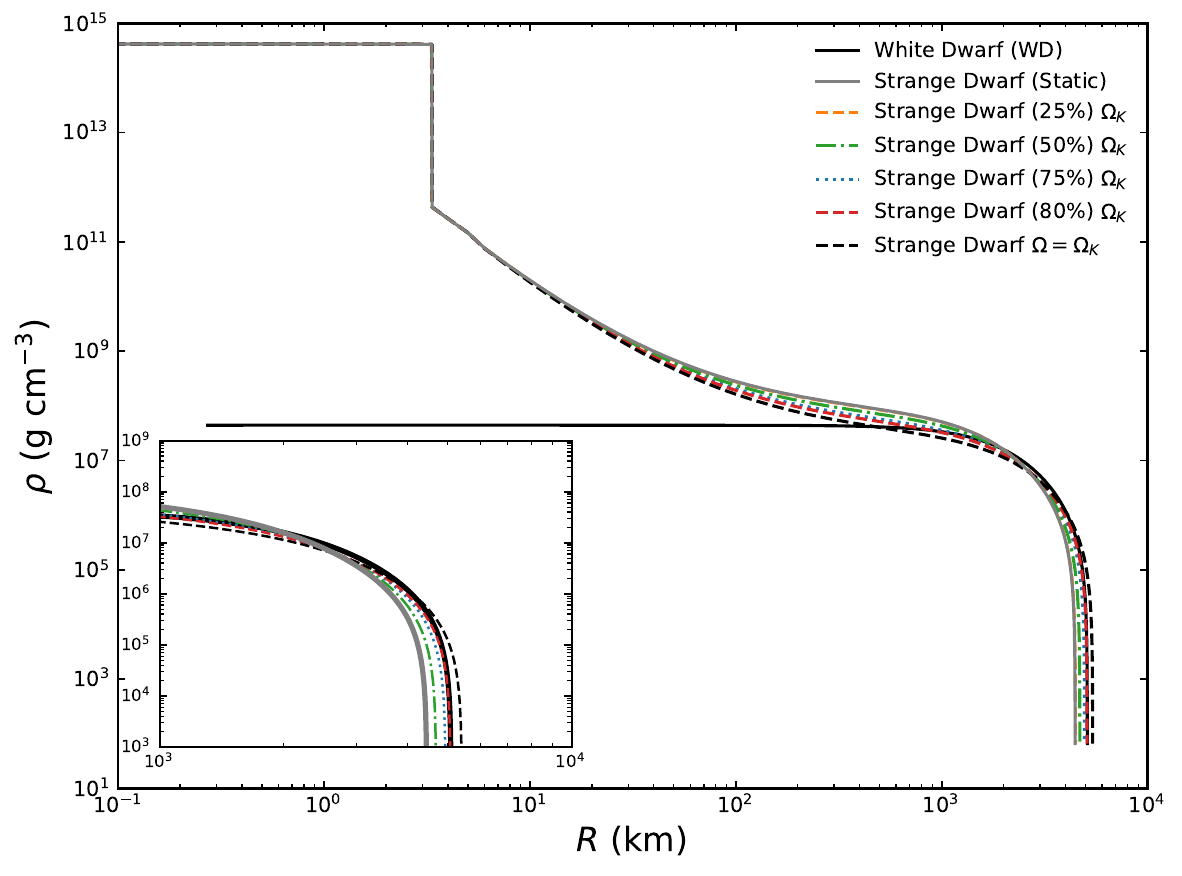}
    \includegraphics[width=0.48\textwidth]{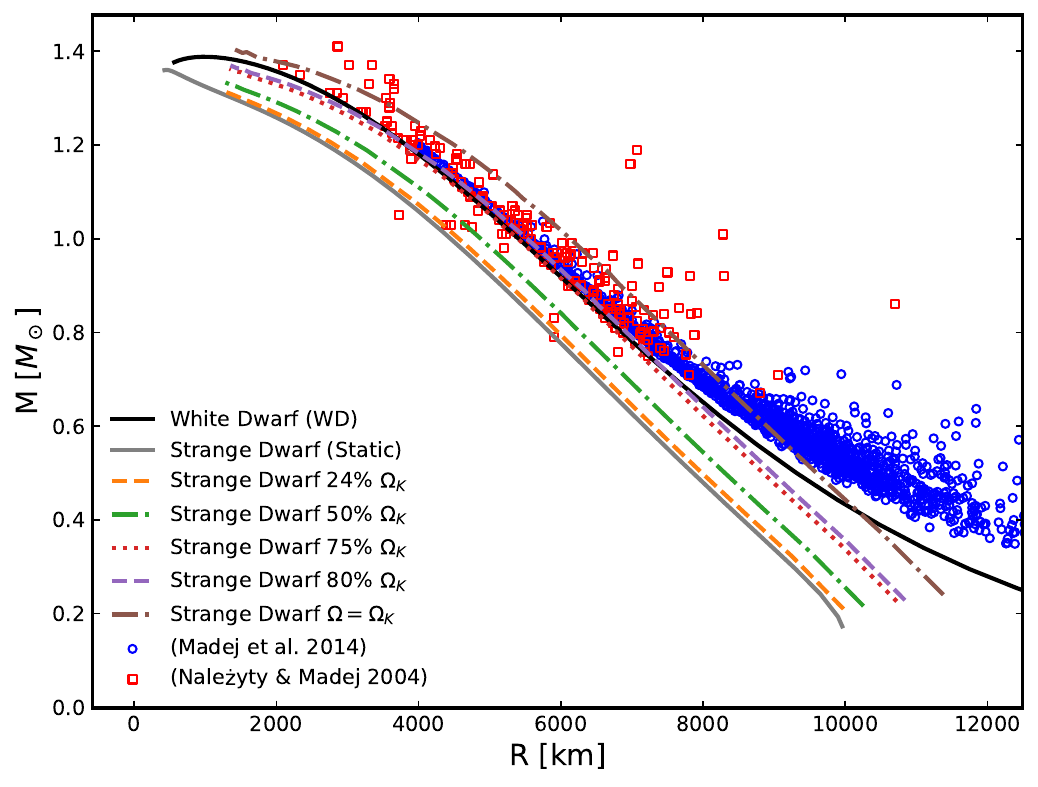}
    \caption{Top: Radial energy-density profiles for a $1\,M_{\odot}$ configuration. The SD profile shows an extended near-constant-density quark core and an abrupt density jump at the core--crust interface (at fixed pressure), whereas the WD profile is smooth. Bottom: $M$--$R$ relations for a pure carbon crust and a quark-matter core, compared with observational data (blue circles from \cite{Madej2004} and red squares from \cite{Nalezyty2004}). The key takeaway is that increasing rotation inflates the SD radius and drives the rotating SD sequences toward the WD locus in the $(M,R)$ plane, motivating a quantitative ``indistinguishability'' criterion discussed in the main text.}
    \label{fig:rhoR}
\end{figure}

Having established the microscopic foundation through the EoS, we now turn to the macroscopic implications of stellar rotation, particularly the role of the Keplerian limit in determining the equilibrium configuration and observable properties of SDs. Figure~\ref{fig:rhoR} presents the energy density and mass profiles as functions of the radial coordinate and total radius in the top and bottom panels, respectively, for a static WD, a static SD, and uniformly rotating SDs at various fractions of the Keplerian frequency \(\Omega_K\). In the top panel, the static WD exhibits an energy density that remains nearly constant with increasing radius until it approaches the stellar surface, where it then decreases monotonically. In contrast, the SD maintains an almost uniform energy density throughout its quark matter core. At the core envelope interface, the density drops abruptly at fixed fluid pressure, after which it decreases monotonically with radius through the crystalline envelope. In the bottom panel, the impact of rotation is clearly visible. These theoretical results are shown alongside observational data extracted from the catalogs listed in \cite{Madej2004} and \cite{Nalezyty2004}, plotted as blue circles and red squares, respectively.

To move beyond a purely visual comparison in the $(M,R)$ plane, we quantify practical indistinguishability between SDs and WDs at fixed mass through the fractional radius difference
\begin{equation}
\Delta R(M,\Omega)\equiv \frac{|R_{\rm SD}(M,\Omega)-R_{\rm WD}(M,0)|}{R_{\rm WD}(M,0)}.\label{eq:deltaR_indist}
\end{equation}
Here, \(R_{\rm SD}(M,\Omega)\) is taken from the rotating SD sequences, while \(R_{\rm WD}(M,0)\) corresponds to the non-rotating WD sequence at the same mass. This expression measures, in percentage terms, how close a rotating SD radius is to the non-rotating WD radius. Table~\ref{tab:deltaR_indist} lists $\Delta R$ for representative masses and rotation rates, providing a direct mass-dependent diagnostic of SD/WD overlap and highlighting the configurations with the smallest separation.

\begin{table*}[ht]
\centering
\caption{Stellar mass, the rotation rate, the SD radius, the corresponding WD radius, the absolute radius difference, and the fractional radius difference, which is defined in Eq.~\eqref{eq:deltaR_indist}.}
\label{tab:deltaR_indist}
\begin{tabular}{cccccc}
\hline
$M/M_\odot$ & $\Omega$ & $R_{\rm SD}$ (km) & $R_{\rm WD}$ (km) & $|R_{\rm SD}-R_{\rm WD}|$ (km) & $\Delta R$ \\
\hline
0.6 & $0.25\,\Omega_K$ & 7301.83 & 8423.79 & 1121.95 & 0.1332 \\
0.6 & $0.50\,\Omega_K$ & 7619.52 & 8423.79 & 804.27  & 0.0955 \\
0.6 & $0.75\,\Omega_K$ & 8150.84 & 8423.79 & 272.94  & 0.0324 \\
0.6 & $0.80\,\Omega_K$ & 8283.44 & 8423.79 & 140.35  & 0.0167 \\
0.6 & $1.00\,\Omega_K$ & 8899.69 & 8423.79 & 475.91  & 0.0565 \\
1.0 & $0.25\,\Omega_K$ & 4565.94 & 5407.96 & 842.02  & 0.1557 \\
1.0 & $0.50\,\Omega_K$ & 4873.43 & 5407.96 & 534.54  & 0.0988 \\
1.0 & $0.75\,\Omega_K$ & 5394.22 & 5407.96 & 13.74   & 0.0025 \\
1.0 & $0.80\,\Omega_K$ & 5524.34 & 5407.96 & 116.37  & 0.0215 \\
1.0 & $1.00\,\Omega_K$ & 6132.75 & 5407.96 & 724.79  & 0.1340 \\
\hline
\end{tabular}
\end{table*}

The minimum values of $\Delta R$ appearing in Table~\ref{tab:deltaR_indist} occur at \(M=0.6\,M_\odot\) for \(\Omega\approx0.80\,\Omega_K\) and at \(M=1.0\,M_\odot\) for \(\Omega\approx0.75\,\Omega_K\) , with \(\Delta R\sim10^{-2}\) and \(\sim10^{-3}\), respectively. This provides a quantitative criterion to determine the mass threshold at which SDs and WDs become practically indistinguishable.

As the rotation rate increases, the stellar radius of an SD grows and approaches the characteristic radius of a WD. This comparison suggests that compact objects containing quark matter may be observationally misclassified as WDs, given that measurements are typically limited to global quantities such as mass and radius, highlighting the urgent need for observational diagnostics capable of probing their innermost composition. Therefore, complementary diagnostics such as tidal deformability or continuous gravitational wave emission may be essential for unambiguously identifying rotating SDs \cite{Glendenning1995May,Perot2023May}

In addition, because the mass radius curves exhibit very little sensitivity to reasonable variations in the quark matter sector as discussed in Subsection \ref{Eos} and illustrated in Figs. \ref{fig:eos_multi_B} and \ref{fig:mass_radiuas_bag} varying the bag constant within the range \({\cal B}^{1/4} = 135, 145, 160~\mathrm{MeV}\) modifies the maximum mass and corresponding radius by only a few percent (see, also, Table~\ref{tab:Bscan}). Thus, the overall morphology of the \(M\)–\(R\) curves, as well as their overlap with the WD domain, remains essentially unchanged. Thus, the observational indistinguishability between rotating SDs and massive WDs does not depend on a finely tuned choice of the bag constant.

\begin{figure}[h]
    \centering
    \includegraphics[width=0.48\textwidth]{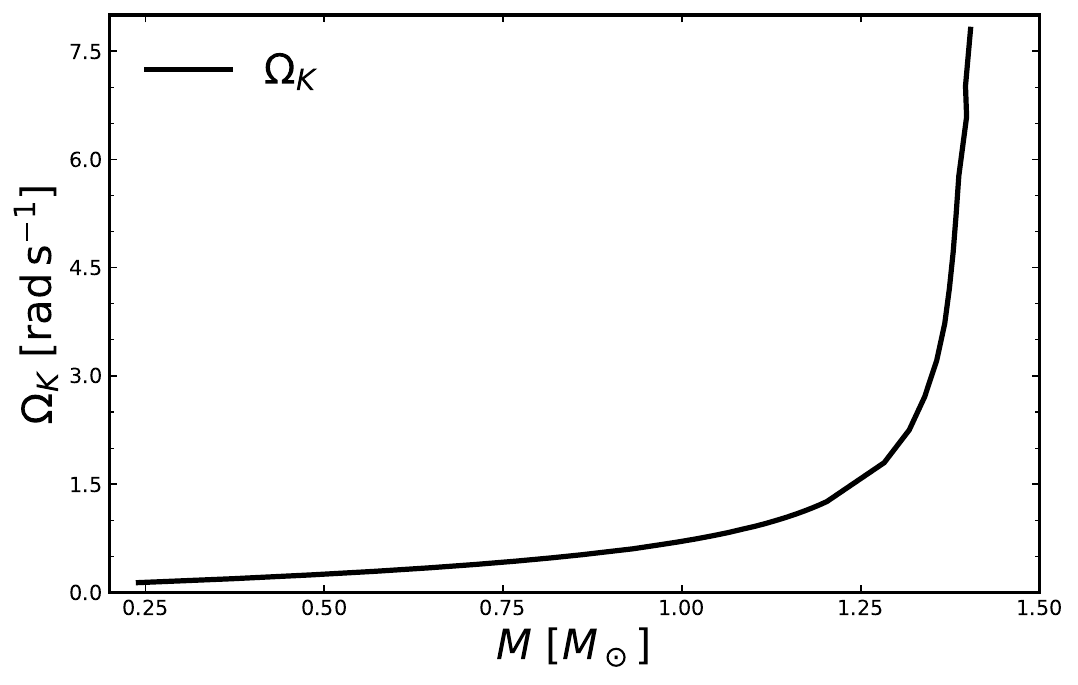}
    \caption{Kepler (mass-shedding) angular frequency $\Omega_{K}(M)$ for the SD sequences, shown here as a reference spin scale. Within the Hartle--Thorne approach, $\Omega_K$ is used to normalize the rotation rate; statements very near $\Omega\sim\Omega_K$ should be interpreted conservatively given the slow-rotation truncation.}
    \label{fig:periods}
\end{figure}

\section{Conclusion}\label{sec:conclusion}

The present investigation analyzes how the presence of a quark matter core would imprint itself on the global properties of compact stars that, at first glance, resemble ordinary WDs. The stellar fluid is modeled with a crust composed of a degenerate electron gas and a core of quark matter. By solving the TOV equations in conjunction with the Hartle-Thorne general relativistic formalism with uniform rotation, we construct sequences of SDs ranging from the non-rotating limit up to their individual Kepler frequencies, where the mass shedding occurs.

For this model, three fundamental conclusions can be drawn. (1) The associated decrease in central density implies that, for \(R\gtrsim 10^{3}\,\mathrm{km}\), the pressure gradient is dominated by the electron gas, while the quark-matter core although still present becomes hydrostatically negligible; since, using this model, radii similar to those of white dwarfs are found. This convergence is illustrated by the radial energy density profiles shown in Fig.~\ref{fig:rhoR}. (2) Rotation shifts the mass-radius sequence curves of the SD towards those of the WD branch. Using the quantitative criterion defined in equation ~\eqref{eq:deltaR_indist}, we observe that the strongest overlap depends on both the mass and the value of the rotation. For $M = 0.6\,M_\odot$, the minimum separation of these curves occurs at $\Omega \approx 0.80\,\Omega_K$, with a fractional radius difference $\Delta R \approx 1.7 \times 10^{-2}$, while for $M = 1.0\,M_\odot$ this occurs at $\Omega \approx 0.75\,\Omega_K$ with a value of $\Delta R \approx 2.5 \times 10^{-3}$.
(3) The computed Keplerian curve, \(\Omega_K(M)\), defines a strict centrifugal limit. Nevertheless, for any given mass, there exists a broad sub-Keplerian range in which the mass of an SD can coincide with that of a WD, rendering the two classes observationally indistinguishable when only global parameters such as mass or radius are considered.

The results reported in this article corroborate and significantly extend the seminal conjecture of Alcock, Farhi, and Olinto (1986) that SQM cores may reside within stars exhibiting WD-like radii. In contrast to Chandrasekhar’s classical $1939$ mass limit \cite{chandrabook}, our models show that incorporating rapid rotation and a deconfined quark-matter core increases the effective upper mass of WD-like objects by \(\gtrsim 10\%\) at the Keplerian threshold \(\Omega_K\), while preserving the canonical limit \(\leq 1.4\,M_{\odot}\) for slowly rotating configurations. By employing the Hartle--Thorne formalism within full general relativity rather than relying on a Newtonian centrifugal correction this work extends the pioneering analysis of uniformly rotating dwarfs by Hartle and Thorne $(1968)$ \cite{hartle1968}. Given that Hartle--Thorne is a slow rotation expansion, our results are best interpreted as capturing robust leading-order rotational trends and delineating where SD/WD degeneracy emerges, while the immediate vicinity of mass shedding should be revisited with fully two-dimensional relativistic rotating-star calculations.

Crucially, our results indicate that global observables alone such as mass, radius, luminosity, or spin period are insufficient to reveal the presence of quark matter when \(\Omega/\Omega_K \lesssim 0.8\). This constrains the parameter space in which alternative diagnostics (e.g., asteroseismology, gravitational redshift spectroscopy, or tidal deformability measurements from inspiral events) must be pursued.

It is worth highlighting that the adoption of a carbon crust in the hybrid EoS does not constitute a restrictive assumption. Theoretical estimates show that heavier crusts (O/Ne/Mg) would modify the stellar mass and radius by less than \(\sim 3\%\) \citep{malheiro_relevance_2021}, a level well below the precision of current mass-radius measurements \citep{mathews_analysis_2006}. Consequently, the main conclusion of this work that rotation obscures the presence of a quark core in SDs remains robust against plausible variations in crustal composition. In addition to the crustal composition, we have explicitly tested the dependence of our results on the quark-matter sector by performing a bag constant sensitivity analysis over the range of \(\cal B\). As shown in Sec. \ref{sec:level2} and summarized in Table \ref{tab:Bscan}, the resulting variations in the maximum mass, corresponding radius, and central density remain at the level of only a few percent, preserving the overall morphology of the \(M\)-\(R\) curves. This demonstrates that rapid rotation can effectively mask the presence of a quark core in SDs.

Finally, this study contributes new nuance to ongoing debates concerning the astrophysical census of compact objects. Because sub Keplerian SDs occupy the same region of the \(M\)–\(R\) diagram as WDs, a non-negligible fraction of the current WD catalog classified based on photometry and parallax may, in fact, contain hybrid stars. Our sequences provide concrete diagnostic criteria, particularly deviations in the quadrupole moment or in the low-\(\ell\) \(g\)-mode spectra, which could distinguish between the two populations in future high precision surveys.

\begin{acknowledgments}
\noindent EO acknowledges support from FUNCAP(BP6-0241-00335.01.00/25). JDVA thanks Universidad Privada del Norte and Universidad Nacional Mayor de San Marcos for the financial support - RR Nº$\,005753$-$2021$-R$/$UNMSM under the project number B$21131781$.
\end{acknowledgments}

\appendix
\section{Numerical robustness tests}
\label{app:robustness}

To assess the numerical stability of the rotating sequences, we carried out a dedicated robustness analysis using the same implementation and the same physical input adopted in the main calculations. For each configuration, a baseline was first obtained and then compared with four controlled numerical variations: (i) tighter convergence thresholds, to test sensitivity to stricter stopping criteria; (ii) looser convergence thresholds, to probe the opposite limit and bracket solver dependence; (iii) a finer radial grid, to increase resolution across the core--crust transition region; and (iv) a coarser radial grid, to quantify the impact of reduced spatial resolution. This protocol isolates numerical effects, since no microphysical parameter of the EoS was modified. For each run, we extracted the global observables $M$, $R$, and $J$, and quantified deviations relative to the corresponding baseline.
\begin{subequations}
\begin{align}
\frac{\Delta M}{M} &= \frac{|M-M_{\rm base}|}{M_{\rm base}},\\
\frac{\Delta R}{R} &= \frac{|R-R_{\rm base}|}{R_{\rm base}},\\
\frac{\Delta J}{J} &= \frac{|J-J_{\rm base}|}{J_{\rm base}}.
\end{align}
\end{subequations}
The test was repeated for the rotating datasets used in this work (including the sequences labeled by fractions of $\Omega_K$). The aggregate ranges found for each perturbation class are reported in Table~\ref{tab:robustness_ranges}.

\begin{table}[ht]
\centering
\caption{Range of relative variations in global observables with respect to baseline runs, from the numerical robustness campaign.}
\label{tab:robustness_ranges}
\begin{tabular}{lccc}
\hline
Setup variation & $\Delta M/M$ & $\Delta R/R$ & $\Delta J/J$ \\
\hline
Tighter thresholds & $0.021$--$0.122$ & $0.020$--$0.135$ & $0.106$--$2.984$ \\
Looser thresholds  & $0.000$--$0.046$ & $0.000$--$0.045$ & $0.000$--$0.200$ \\
Finer grid         & $0.016$--$0.041$ & $0.033$--$0.057$ & $0.042$--$0.062$ \\
Coarser grid       & $0.030$--$0.076$ & $0.063$--$0.111$ & $0.056$--$0.118$ \\
\hline
\end{tabular}
\end{table}

Across all tested numerical setups, the qualitative behavior reported in the main text is preserved: increasing rotation systematically enlarges the stellar radius and drives the SD sequence toward the WD region in the $(M,R)$ plane. This confirms that the SD--WD overlap discussed in this work is not tied to a single numerical realization. At the same time, the robustness analysis provides a quantitative estimate of the numerical dispersion in global observables. The largest deviations are obtained in the tight-threshold test for the lowest-spin sequence, reflecting the sensitivity of the numerical integration across the sharp core–crust interface. In all cases, however, the qualitative rotational trend and the SD–WD overlap in the (M, R) plane remain unchanged.. The complete set of run-by-run results is presented in Table~\ref{tab:robustness_full}.

\begin{table*}[ht]
\centering
\caption{Complete numerical robustness results for rotating SD sequences. In all setups, the baseline model is approximately at $M\sim 1\,M_\odot$. Relative deviations are computed with respect to the baseline of each sequence.}
\label{tab:robustness_full}
\begin{tabular}{llcccccc}
\hline
Sequence & Setup & $M\,(M_\odot)$ & $R$ (km) & $J/(G M_\odot^2)$ & $\Delta M/M$ & $\Delta R/R$ & $\Delta J/J$ \\
\hline
$0.25\,\Omega_K$ & Baseline            & 1.0084 & 4503.931 & 1.61448 & 0.00000 & 0.00000 & 0.00000 \\
$0.25\,\Omega_K$ & Tighter thresholds  & 1.1314 & 5110.244 & 6.43261 & 0.12198 & 0.13462 & 2.98432 \\
$0.25\,\Omega_K$ & Looser thresholds   & 1.0084 & 4503.931 & 1.61448 & 0.00000 & 0.00000 & 0.00000 \\
$0.25\,\Omega_K$ & Finer grid          & 1.0257 & 4357.548 & 1.52333 & 0.01716 & 0.03250 & 0.05646 \\
$0.25\,\Omega_K$ & Coarser grid        & 0.9742 & 4785.934 & 1.78910 & 0.03392 & 0.06261 & 0.10816 \\
\hline
$0.50\,\Omega_K$ & Baseline            & 1.0024 & 4855.823 & 3.21097 & 0.00000 & 0.00000 & 0.00000 \\
$0.50\,\Omega_K$ & Tighter thresholds  & 1.0962 & 5336.452 & 6.27716 & 0.09358 & 0.09898 & 0.95491 \\
$0.50\,\Omega_K$ & Looser thresholds   & 1.0024 & 4855.823 & 3.21097 & 0.00000 & 0.00000 & 0.00000 \\
$0.50\,\Omega_K$ & Finer grid          & 1.0200 & 4669.341 & 3.02233 & 0.01756 & 0.03840 & 0.05875 \\
$0.50\,\Omega_K$ & Coarser grid        & 0.9679 & 5227.827 & 3.57308 & 0.03442 & 0.07661 & 0.11277 \\
\hline
$0.75\,\Omega_K$ & Baseline            & 1.0073 & 5336.877 & 4.76630 & 0.00000 & 0.00000 & 0.00000 \\
$0.75\,\Omega_K$ & Tighter thresholds  & 1.0544 & 5590.844 & 6.04499 & 0.04676 & 0.04759 & 0.26828 \\
$0.75\,\Omega_K$ & Looser thresholds   & 1.0073 & 5336.877 & 4.76630 & 0.00000 & 0.00000 & 0.00000 \\
$0.75\,\Omega_K$ & Finer grid          & 1.0235 & 5083.536 & 4.47600 & 0.01608 & 0.04747 & 0.06091 \\
$0.75\,\Omega_K$ & Coarser grid        & 0.9767 & 5866.485 & 5.32160 & 0.03038 & 0.09924 & 0.11651 \\
\hline
$0.80\,\Omega_K$ & Baseline            & 0.9968 & 5544.305 & 5.03996 & 0.00000 & 0.00000 & 0.00000 \\
$0.80\,\Omega_K$ & Tighter thresholds  & 1.0304 & 5731.546 & 5.89010 & 0.03371 & 0.03377 & 0.16868 \\
$0.80\,\Omega_K$ & Looser thresholds   & 0.9968 & 5544.305 & 5.03996 & 0.00000 & 0.00000 & 0.00000 \\
$0.80\,\Omega_K$ & Finer grid          & 1.0137 & 5269.451 & 4.72838 & 0.01695 & 0.04957 & 0.06182 \\
$0.80\,\Omega_K$ & Coarser grid        & 0.9650 & 6159.333 & 5.63423 & 0.03190 & 0.11093 & 0.11791 \\
\hline
SD rotating ($\sim\Omega_K$) & Baseline            & 0.9932 & 6179.590 & 6.12796 & 0.00000 & 0.00000 & 0.00000 \\
SD rotating ($\sim\Omega_K$) & Tighter thresholds  & 0.9726 & 6055.023 & 5.47566 & 0.02074 & 0.02016 & 0.10645 \\
SD rotating ($\sim\Omega_K$) & Looser thresholds   & 0.9475 & 5902.839 & 4.90132 & 0.04601 & 0.04478 & 0.20017 \\
SD rotating ($\sim\Omega_K$) & Finer grid          & 1.0343 & 5829.149 & 6.38540 & 0.04138 & 0.05671 & 0.04201 \\
SD rotating ($\sim\Omega_K$) & Coarser grid        & 0.9178 & 6724.250 & 5.78642 & 0.07592 & 0.08814 & 0.05573 \\
\hline
\end{tabular}
\end{table*}

\end{document}